\documentclass[%
 reprint,
superscriptaddress,
 amsmath,amssymb,
 aps,
prb,
]{revtex4-1}

\usepackage{graphicx}
\usepackage{dcolumn}
\usepackage{bm}
\usepackage{amsmath,amssymb}
\usepackage{graphicx}
\usepackage{color}

\usepackage{hyperref}

 \newcommand{\un}[1]{\;\mathrm{#1}}

\begin{document}
\preprint{APS/123-QED}
\title{Laser-induced Translative Hydrodynamic Mass Snapshots: mapping at nanoscale}



\author{X. W. Wang}
\affiliation{Center for Micro-Photonics, Swinburne University of Technology, John st., Hawthorn 3122, Victoria, Australia}

\author{A. A. Kuchmizhak}
\email{alex.iacp.dvo@mail.ru}
\affiliation{Center for Micro-Photonics, Swinburne University of Technology, John st., Hawthorn 3122, Victoria, Australia}
\affiliation{School of Natural Sciences, Far Eastern Federal University, Vladivostok, Russia}
\affiliation{Institute of Automation and Control Processes, Far Eastern Branch, Russian Academy of Science, Vladivostok 690041, Russia}

\author{X. Li}
\affiliation{Center for Micro-Photonics, Swinburne University of Technology, John st., Hawthorn 3122, Victoria, Australia}

\author{S. Juodkazis}
\affiliation{Center for Micro-Photonics, Swinburne University of Technology, John st., Hawthorn 3122, Victoria, Australia}
\affiliation{Melbourne Centre for Nanofabrication, ANFF, 151 Wellington Road, Clayton, VIC 3168, Australia}

\author{O. B. Vitrik}
\affiliation{School of Natural Sciences, Far Eastern Federal University, Vladivostok, Russia}
\affiliation{Institute of Automation and Control Processes, Far Eastern Branch, Russian Academy of Science, Vladivostok 690041, Russia}

\author{Yu. N. Kulchin}
\affiliation{Institute of Automation and Control Processes, Far Eastern Branch, Russian Academy of Science, Vladivostok 690041, Russia}

\author{V.~V.~Zhakhovsky}
\affiliation{Dukhov Research Institute of Automatics, ROSATOM, Moscow 127055, Russia}
\affiliation{Landau Institute for Theoretical Physics, Russian Academy of Sciences, Chernogolovka 142432, Russia}

\author{P. A. Danilov}
\affiliation{Institute of Automation and Control Processes, Far Eastern Branch, Russian Academy of Science, Vladivostok 690041, Russia}
\affiliation{Lebedev Physical Institute, Russian Academy of Sciences, Moscow 119991, Russia}

\author{A. A. Ionin}
\affiliation{Lebedev Physical Institute, Russian Academy of Sciences, Moscow 119991, Russia}

\author{S. I. Kudryashov}
\affiliation{Institute of Automation and Control Processes, Far Eastern Branch, Russian Academy of Science, Vladivostok 690041, Russia}
\affiliation{Lebedev Physical Institute, Russian Academy of Sciences, Moscow 119991, Russia}
\affiliation{ITMO University, St. Peterburg 197101, Russia}
\affiliation{National Research Nuclear University MEPhI (Moscow Engineering Physics Institute), Moscow 115409, Russia}

\author{A. A. Rudenko}
\affiliation{Lebedev Physical Institute, Russian Academy of Sciences, Moscow 119991, Russia}

\author{N.~A.~Inogamov}
\affiliation{Landau Institute for Theoretical Physics, Russian Academy of Sciences, Chernogolovka 142432, Russia}
\affiliation{Dukhov Research Institute of Automatics, ROSATOM, Moscow 127055, Russia}


\date{\today}
\begin{abstract}
Nanoscale thermally assisted hydrodynamic melt perturbations induced by ultrafast laser energy deposition in noble-metal films produce irreversible nanoscale translative mass redistributions and results in formation of radially-symmetric frozen  surface structures. We demonstrate that the final three-dimensional (3D) shape of the surface structures formed after re-solidification of the molten part of the film is governed by incident laser fluence and, more importantly, predicted theoretically via molecular dynamics modeling. Considering the underlying physical processes associated with laser-induced energy absorption, electron-ion energy exchange, acoustic relaxation and hydrodynamic flows, the theoretical approach separating ``slow'' and ``fast'' physical processes and combining hybrid analytical two-temperature calculations, scalable molecular-dynamics simulations, and a semi-analytical thin-shell model was shown to provide accurate prediction of the final nanoscale solidified morphologies, fully consistent with direct electron-microscopy visualization of nanoscale focused ion-beam cuts of the surface structures produced at different incident laser fluences. Finally, these results are in reasonable quantitative agreement with mass distribution profiles across the surface nanostructures, provided by their noninvasive and non-destructive nanoscale characterization based on energy-dispersive x-ray fluorescence spectroscopy, operating at variable electron-beam energies.

\end{abstract}

\maketitle

\section{\label{sec:Intro} Introduction}
Precise high numerical-aperture (NA) nanoscale laser ablation of thin films, using short 
and ultrashort 
laser pulses, is a promising physical effect 
crucial for development of emerging technologies in processing thin-film transistors~\cite{Wang:2007}, scribing thin-film solar cells~\cite{Stefan:2008}, ablative fabrication and light-induced forward transfer (LIFT)-printing of advanced plasmonic and dielectric nanophotonic metasurfaces and
circuits~\cite{Vorobyev13,Urs:2014,Visser:2015,Dmitriev:2016,Kuchmizhak:Nanoscale2016,Wang:2017,Kuznetsov:2011,Zenou2015,Fang17,Makarov17}. In comparison to short laser pulses, the ultrashort ones are broadly used during such precise ablation, holding a promise of providing an ultimate spatial resolution, despite the underlying more intense
response of electron subsystem to laser exposure with dramatically higher electron/lattice temperature (if $\tau_L<1$ ps) and pressure gradients, resulting in intense nanoscale hydrodynamic flows and ultrafast quenching of corresponding transient melt configurations - e.g., a nanodroplet, separating from a nanojet \cite{Unger:2012}. Commonly, such nano- and microscale structures resolidified on thin supported metallic films, are qualitatively or semi-quantitatively visualized by top- or side-view scanning electron microscopy (SEM), with just a few sketchy studies revealing their internal structure for some textures without quantitative acquisition of their parameters and without visualization of "hidden" sub-surface features (e.g., counter-jets, cavities)~\cite{Kuznetsov:2012,J:2014,Nakata2013,Yuan2013,Wu:2015,Ivanov2015}. On the contrary, molecular dynamics (MD) and hydrodynamics (HD) simulations provide \cite{Ivanov2015,Wu2016,Inogamov:2016:Nanoscale,Itina2007} detailed envision of the underlying spatiotemporal dynamics of the nanoscale hydrodynamic flows as well as describe the corresponding physical mechanisms, however, don't have a starting point for future important \emph{quantitative} predictions and optimization of experimental parameters, required for ultimate resolution as well as no firm experimental background for comparative justification of their results.

\indent In this study, SEM inspection of nanoscale focused-ion beam (FIB) cuts and energy-dispersive X-ray fluorescence (EDX) nanoscale profiling of individual radially-symmetric topological surface structures - separate nano-bumps and jets on nano-bumps, produced by single-shot ablation of 50-nm-thick glass-supported gold films by tightly focused fs-laser pulses of variable energies were conducted to reveal quenched configurations of melt, yielding from irreversible nanoscale translative hydrodynamic flows and nanoscale heat conduction. By taking into account all underlying physical processes associated with laser-induced energy absorption, electron-ion energy exchange, acoustic relaxation and hydrodynamic flows, the theoretical approach based on separation of "slow" and "fast" physical processes and combining hybrid analytical two-temperature calculations, scalable molecular-dynamics simulations and semi-analytical thin-shell model was shown to provide reasonable accuracy in prediction of such nanoscale resolidified morphologies.

\section{\label{sec:SEM} Experimental study of nanoscale translative mass redistributions}

To reveal for the first time the initial stage of metal film blistering from its substrate under the tightly-focused fs-laser pulse irradiation, as well as to follow the subsequent evolution of the molten material -- nanoscale hydrodynamic flows and quenching of transient nanofeatures -- from the parabola-shaped nanobumps to the small nanojets, we have fabricated cross-sectional cuts using $\un{Ga^+}$-ion FIB milling (Raith IonLINE). To do this, an e-beam evaporated 50-nm thick Au film, covering a silica glass substrate, was first patterned with well-ordered arrays of different surface structures -- smaller and larger nanobumps; with small nanojets atop for the larger fluences (Fig. 1(a-e)). The spatial period of arrays was set $2\un{\mu m}$.

\begin{figure}[h!]
\includegraphics[width=0.9\columnwidth]{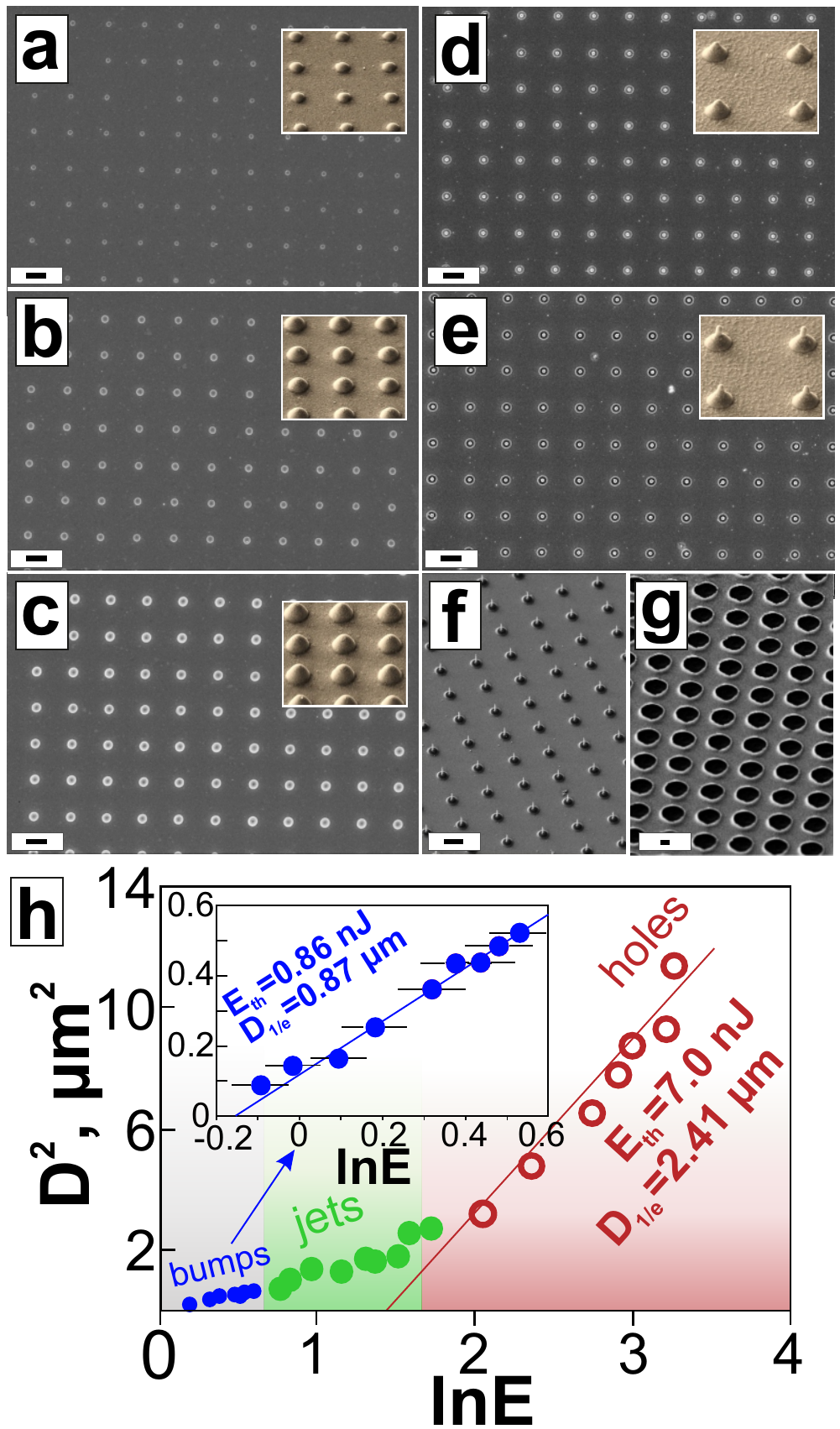}
\caption{(Color online) (a-e) SEM images of the nanostructure arrays fabricated at increasing pulse energies in units of nJ: (a) 0.9, (b) 1.2, (c) 1.38, (d) 1.62, and (e) 1.7 nJ; the insets present side-view (view angle of $45^\circ)$ SEM images of the corresponding structures. (f,g) Side-view (view angle of $45^\circ)$ SEM images of nanojet and microhole arrays fabricated at pulse energies: (f) 3.3 and (g) 40.5 nJ, respectively; The scale bars for these images correspond to $1 \un{\mu m}.$ Frame (h) shows squared outer diameter $D^2$ of the nanobumps, nanojets and through holes versus natural logarithm of the applied pulse energy $\ln E$ $(E$ in nJ). The inset illustrates the magnified left-most part of this dependence. The slope of the fitting lines for bumps and through holes indicates the characteristic energy deposition diameter, while the intersection with the $x$-axis -- the threshold pulse energies required to produce the bump and hole structures, respectively.}
\label{fig:1}
\end{figure}

The small structures are formed in the fluence range below the thresholds for nanojet and hole appearance (highlighted by green and red colors in Fig.~\ref{fig:1}(h), respectively). 
Studying of the small structures can shed light on the reasons of material redistribution during their initial formation stage, because the main governing
processes, including melting, material flow, and recrystallization, are triggered at such timescale. Additionally, these experimental results can be used to justify for the first time the previously suggested theoretical models \cite{Inogamov:2016:Nanoscale,Meshcheryakov13,Ivanov13,Meshcheryakov06}.

\begin{figure*}[t]
\includegraphics[width=0.85\textwidth]{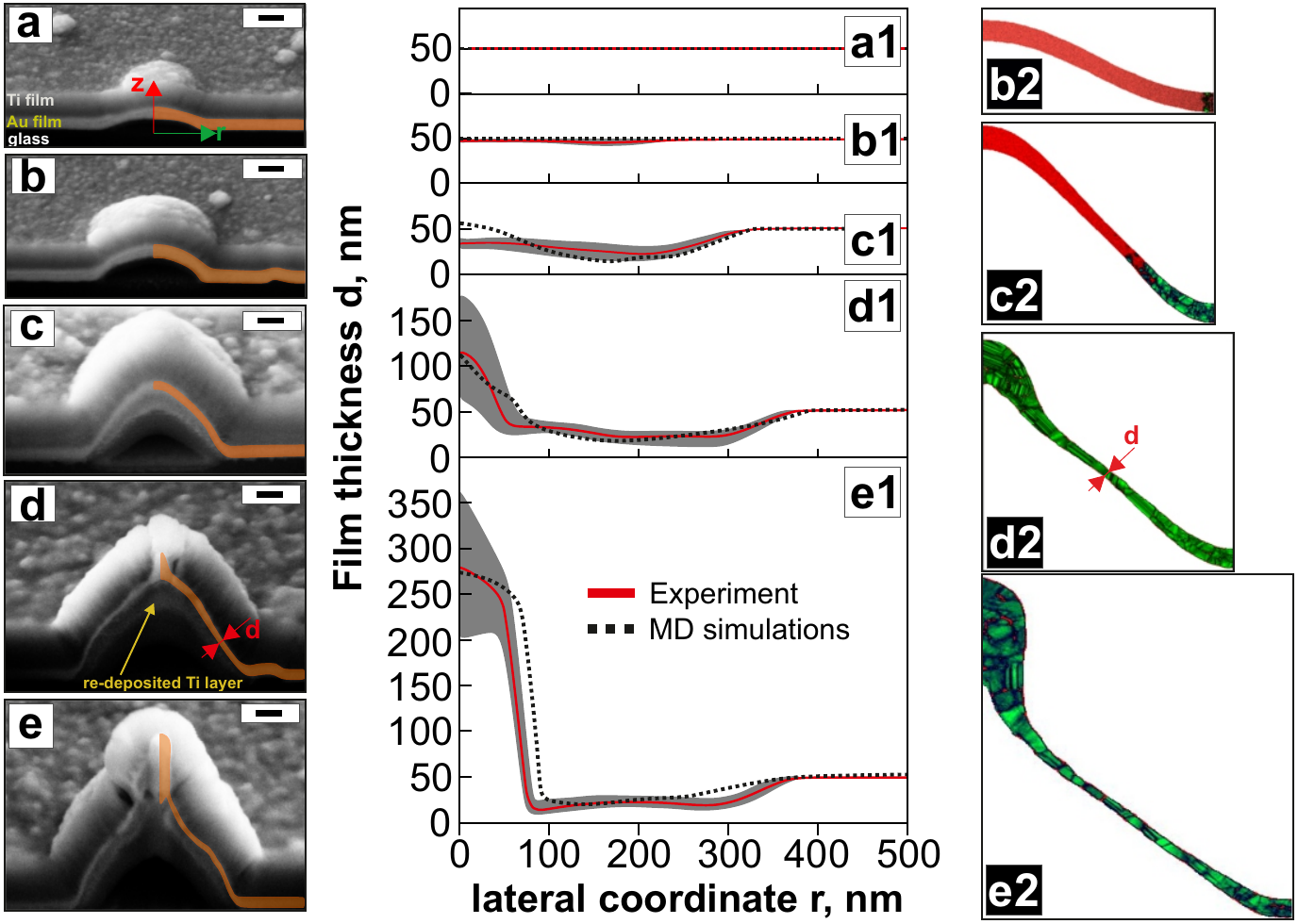}
\caption{(Color online) (a-e) Side-view SEM images (view angle of $45^\circ)$ of FIB cross-sectional cuts of single surface structures
 produced at the increasing pulse energy and previously shown in figure 1 (a-e), respectively. Orange areas in each image highlight the Au film.
 The scale bars for these images correspond to 100 nm. The darker grey layer under the Au film, appearing in the images (c-e),
 is attributed to the re-deposition of Ti during the FIB milling, which was confirmed by the corresponding EDX measurements;
 (a1-e1) Film thickness $d$ experimentally measured over 10 similar cuts of the 10 similar surface structures
 produced at the same irradiation conditions (red curves), and recalculated via corresponding MD simulations (black dotted curve)
 as a function of the lateral $r$-coordinate (cylindrical radius) defined in Fig. 2(a). Film thickness $d$ is measured and recalculated in the directions
  normal to the local section of the film (details are given in inset in Fig.~3(b) below). Grey areas indicate the error bar;
  (b2-e2) molecular-dynamics (MD) simulations showing the modification of the metal shell with absorbed energy. Red and green colors divide the molten and solid states of the film.
} \label{fig:2}
\end{figure*}

For laser patterning, second-harmonic ($\lambda = 515$nm), 230-fs laser pulses from a Yb:KGW laser system (PHAROS, Light Conversion Ltd.) were focused into a sub-micrometer spot using a dry objective lens with NA = 0.5 (Mitutoyo M Plan Apo NIR HR). Each surface structure presented in the Fig.1(a-e) was produced by single-pulse irradiation, keeping the constant pulse energy $E$ for each single array, while increasing stepwise the applied pulse energy from 0.9 to 1.7~nJ/pulse for subsequently patterned arrays (Fig.~1(a-e)).

After such laser fabrication, the sample, containing nanostructures, was overcoated with a protective titanium
(Ti) film. The Ti coating provides a good contrast of FIB cuts of the laser-fabricated nano- and
microstructures during their SEM visualization. For small parabola-shaped microbumps (Fig.1(a,b)), the 100-nm
thick Ti film was post-deposited, while for taller structures a twice thicker protective layer was used to
ensure their complete coverage. The succeeding FIB milling of the sample was performed at the 30-kV
acceleration voltage and the relatively small beam current of 50~pA to produce smooth walls. For each type of
nano- or micro-structure, at least 10 similar cuts were prepared to provide statistical significance and to
reveal small fluctuations due to instability of the laser pulse energy, as well as stochastic inhomogeneities
of the metal film/glass substrate. Finally, the fabricated FIB-cuts were visualized, using a field-emission SEM module of the e-beam lithography writer Raith 150-TWO.

By measuring the dependence of squared diameters $D^{2}$ of the through-holes produced in the Au film vs. natural logarithm of incident pulse energy, $\ln(E),$ the corresponding threshold pulse energy of $E_{th,hole}= 7.0 \pm 0.7$ nJ (red dots in Fig.~1(f)) was estimated, with its slope indicating the characteristic energy deposition diameter $D_{1/e,hole}$ of $2.4 \un{\mu m}$; $E_{th,hole}$ is the point between the green (jet) and red (hole) ranges on the axis $\ln E$ in Fig. 1(f). This gives the deposited threshold fluence $F_{th,hole} = A\cdot E_{th,hole} (\pi R^2_{1/e})^{-1} = 0.046 \pm 0.006 \un{J/cm^2}$ for the 50-nm-thick Au film absorbance $A \approx 0.3$, which is in good agreement with previously reported values; see also Fig.~4 below. Similar dependences measured for nanobumps and jets (blue and green dots in Fig.~1(f)) indicate practically the same deposited fluence $F_{th,bump} = 0.042 \pm 0.006\un{J/cm^2}$ at the threshold pulse energy $E_{th,bump} = 0.86 \pm 0.09$ nJ and almost three-fold smaller energy deposition radius $D_{1/e,bump} = 0.87 \pm 0.09 \un{\mu m}.$ Obviously, the different energy deposition scales - $D_{1/e,bump} \approx 0.9 \un{\mu m}$ and $D_{1/e,hole} \approx 2.4 \un{\mu m}$ -- indicate, for the same focusing conditions, the corresponding different temporal scales of their formation, enabling for the laser-deposited energy in the thin film to be transported laterally from the diffraction-limited focal spot (diameter $D_{1/e,foc} \approx 0.5 \un{\mu m}$ for the 0.5-NA dry objective lens) via electron heat conduction. Specifically, in the cases of bumps and holes $t_{bump} \approx (D_{1/e,bump}^2 - D_{1/e,foc}^2)/4\chi \approx 1$ ns and $t_{hole} \approx (D_{1/e,hole}^2 - D_{1/e,foc}^2)/4\chi \approx 10$ ns, respectively, for the thermal diffusivity coefficient of solid gold $\chi \approx 1.2 \un{cm^2/s},$ being comparable to previous similar estimates.~\cite{Kulchin2014,Danilov:2014}

Series of side-view SEM images (Fig.~2(a-e)) demonstrate the central cross-section cuts of the different structures produced on the surface of the Au film at the increasing pulse energy (each presented image illustrates the cut of one of the surface structure previously shown in Fig. 1(a-e)). The detailed analysis of the cuts indicate that for parabolically-shaped nanobumps produced at the pulse energies slightly above the measured threshold for the nanobump formation (E${}_{th }$$>$ 0.87 nJ), the thickness of the Au film remains almost unchanged with some negligible material redistribution, occurring only near the  edges of the nanobump (Fig.~2(a1,b1)). For the increasing incident pulse energy (Fig.~2(c,d)),  significant thinning of the film is observed (in some specific points near the nanobump edge the film becomes twice thinner) and is associated with the corresponding increase of the nanobump height. The subsequent evolution of the nanobump shape - from parabolic to conical one - occurs at the further increase of the incident pulse energy (Fig.~2(d,d1)) accompanied with accumulation of the molten material at the nanobump tip in the form of the 120-nm-high and 100-nm-wide protrusion. The lateral and vertical dimensions of this protrusion continuously increase versus the incident pulse energy (Fig.~2(d-e,d1-e1)), forming a nanojet, while the surrounding nanobump becomes thinner through the jet-directed nanoscale hydrodynamic flow with its minimal experimentally observed thickness, reaching $\approx$ 13~nm near the nanojet edge and $\approx$ 20~nm near the nanobump edge.

Despite its straightforward character, such FIB cutting of surface nanotextures produced on noble-metal films, which are typically weakly resistant to electron- and ion-beam exposures, requires additional sample over-coating by some protective layer, mostly excluding its following practical nanophotonic applications. In this respect, non-destructive and non-invasive experimental methods, enabling quantitative characterization of supported nanoscale structures, are demanding. In this study, we have performed nanoscale cross-sectional SEM-based EDX analysis schematically illustrated in the Fig. 3(a) of the typical structures shown in the Fig. 2(e). Among these surface nanostructures, the microbump, containing the small nanojet atop, provides the maximal metal thickness redistribution - from the 13-nm-thick cap of the bump to a few hundred nm high protrusion - as it was previously confirmed by SEM analysis of the FIB cuts. Additionally, the steep wall of the central protrusion in Fig.~2(e) represents almost a perfect object to estimate the lateral resolution of this SEM-based EDX approach.
\begin{figure}[t]
	\includegraphics[width=0.8\columnwidth]{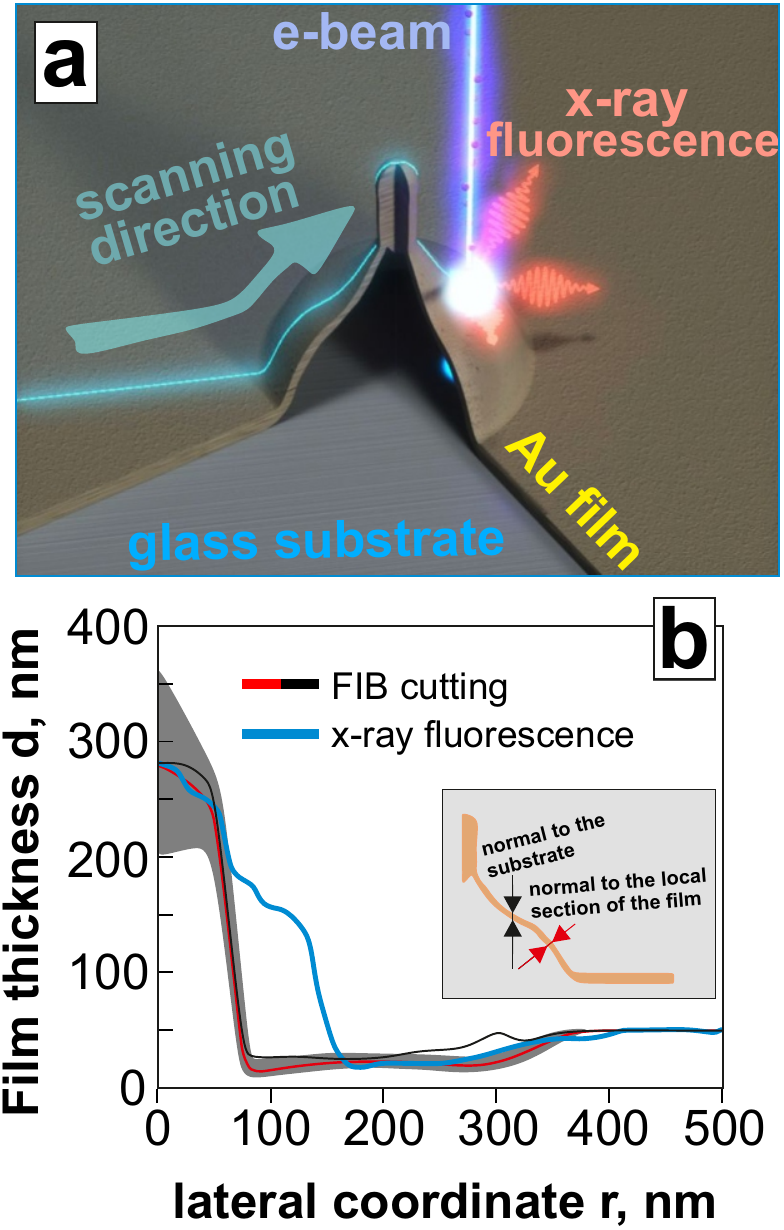}
	\caption{(Color online) (a) Sketch explaining the principle of the EDX profiling across the nanofeature by the e-beam excitation and subsequent detection of the characteristic x-ray fluorescence signal. (b) Radial profiles of film thickness measured using EDX cross-sectional analysis (blue curve) and FIB cutting/SEM visualization recalculated to present the film thickness in the direction normal to the substrate (black curve) and normal to the local section of the film (red curve). The inset shows the difference between the actual thickness along the normal to its local section and the top-view EDX-acquired thickness.}
	\label{fig:3}
\end{figure}

\begin{table*}
  \centering
  \caption{Parameters of MD-MC simulations for two regimes of mass redistribution leading to formation of either bumps or jets. This makes possible to plot the theoretical boundary between these regimes marked as points "d" and "e" on Fig.~4. The corresponding final structures obtained in simulation are shown in Figs. 2(d1,d2,e1,e2).  }
  \label{tab:1}
 \vspace{0.0cm}
\begin{tabular}{|c|c|c|c|c|c|c|c|c|c|}
\hline
 No.       &$N_{atoms}$& $2 r_c,$ nm& $d,$ nm& $\chi.$ cm$\!^2/$s& $\nu_{cm},$ m/s& $\nu_\sigma,$ m/s& $\nu_\chi,$ m/s& $V_\sigma$& $V_\kappa$ \\
\hline
 \emph{d}  & $44\times 10^6$ &  260  & 10   & 0.18  & 200  & 106  & 69   & 0.529     & 0.346      \\
\hline
 \emph{e}  & $6\times 10^6$  &  130  & 5.1  & 0.18  & 385  & 148  & 138  & 0.384     & 0.358      \\
\hline
 \end{tabular}
 \vspace{0.0cm}
 \end{table*}

EDX cross-sectional (radial) elemental micro-analysis was carried out with a field-emission SEM microscope JEOL 7001F using an INCA Energy 350XT spectrometer (Oxford Instruments Analytical), at the electron energies of 10 and 30~keV, while the electron beam current was chosen at the level of 79~pA to avoid 
thermal damage of the nanostructures on the film. The acquisition of the EDX-profiles for the laser-induced surface nanofeatures at the particular electron energies, which provide the penetration depths in the film and the supporting substrate much larger, than the 50-nm film thickness, enables their calibration in terms of thickness, using the reference EDX-signal value for the non-irradiated gold film.

More specifically, the 30-keV electron beam provides too low and noisy EDX-signals from the flattened regions of the thin Au film due to its very deep $(2-3\un{\mu m})$ penetration through the film into the supporting substrate, but the 30-keV beam produces the appropriate EDX-signals across the micron-tall nanofeatures, formed by the laser-driven melt accumulation 
(the central protrusion atop the nanobump). In contrast, the 10-keV electron beam provides reasonable EDX-signals from the gold film and laser-induced features of comparable thickness, but is almost completely absorbed by the micron-tall surface features, excluding calibration of the corresponding EDX-signals. As a result, the acquired cross-sectional EDX-profiles were linked in the transition region from the thinned film within the bump and the tall central protrusion, representing the initial stage of the nanojet formation, with the 30-keV EDX-signal in the center and 10-keV EDX-signal at the periphery.

Figure 3(b) shows the EDX-profiling results for the thickness of the nanojet with the surrounding nanobump (Fig.~2e,e1) in comparison to the side-view SEM measurements of its thickness along the normal to the substrate, using the cross-sectional cuts. Both curves indicate their reasonable agreement within the experimental error bars for all three topographies - the non-irradiated film, nanobump and nanojet. Minor difference in spatial resolution of these approaches indicated by the 100~nm difference of the nanojet radius in Fig.~3, is a broadening artefact of an external electromagnetic noise in the local circuits during the prolongated EDX-profiling procedure (3-5 minutes).

\section{\label{sec:methods} Simulation technique}

In this study we used a comprehensive simulation approach, including the two-temperature electron-ion
hydrodynamics (2T-HD) and the classical molecular dynamics method combined with Monte-Carlo heat transfer by
electrons (MD-MC) \cite{Inogamov:2010:APA}, to represent the next step toward accurate prediction of the
translative flow-assisted redistribution of material after the ultra-short laser exposure. To achieve this
goal, all relevant physical processes underlying the material response to such tightly-localized ultra-fast
energy deposition in a skin layer should be taken into account. However, the simultaneous consideration of
those multiple processes on the micrometer-sized scale during several nanoseconds with our simulation
technique, including billions of interacting atoms and electrons, far exceeds the limits of available
supercomputers. To overcome this limitation, we divide all physical processes into ``fast'' and ``slow''
processes, and apply the 2T-HD or MD-MC model, respectively.


The 2T-HD model, describing the first short evolution stage with its typical duration of the order of acoustic timescale $(\approx 17\un{ps}$ for our 50-nm-thick Au film and longitudinal speed of sound in gold of 3~km/s),
includes one-dimensional two-temperature (2T) hydrodynamic code having full 2T physics: absorption laser energy, 2T electron thermal conduction, energy exchange between electrons and atomic vibrations via their coupling, and 2T equation of state. Thus, the model takes into accounts all fast processes associated with electron pressure, electron conductivity, electron-ion temperature relaxation in gold lasting 7-10 ps,~\cite{Ilnitsky:2016:JPCC}, and separation of the molten film from the substrate with separation velocity $\nu_{cm}(r)$ depending on the local stress $p(r)$ generated by isochoric laser heating of film up to lattice temperatures of few thousand degrees produced by the absorbed fluence $F_{abs}(r)$.


The stress of $p(r) \sim 10$GPa, generated within the fist stage, is responsible for separation of film from
the substrate. The 2T-HD model provides the separation velocities $\nu_{cm}\approx (Z_{s}/Z_{f}) (p/B)c_s$
ranging from 0 to $\approx 60$ m/s, where $c_{s}$ is the longitudinal speed of sound in gold, $B$ is the bulk
modulus of gold, $Z_s$, $Z_f$ are acoustic impedances of the substrate and the film, respectively, see details
in the Supplemental material. In the Supplemental material we discuss comparison between velocities $\nu_{cm}$ in simulations and
experiments.


The second, "slow" evolution stage starts after electron-ion thermal equilibration and completes after
recrystallization of surface melt, which requires tens of nanoseconds typically. To simulate the slow processes
initiated within the fast stage the combined MD-MC approach is utilized. The profiles of temperature and
initial vertical velocity of film obtained after the first stage in series of the 2T-HD modeling with different
absorbed fluences are used as the input data for starting three-dimensional MD-MC simulations.

Then, the embedded atom method (EAM) interatomic potential for gold was used \cite{Zhakhovskii:2009}, reproducing well the stress-strain and thermal characteristics of gold. The potential gives the density of melt about $16.9\un{g/cc}$ and the melting point of about 1330 K, which is close to the experimental $T_m=1337$K. But the calculated surface tension of $0.546\un{J/m^2}$ is notably lower than the experimental $\sigma=1.125\un{J/m^2}$ at the melting temperature, see discussion in Appendix of \cite{Inogamov:2015:JETP}.

The simulation parameters of two selected MD-MC runs are presented in Table 1. A thin film with thickness $d$ is placed in a square simulation box with dimensions $L_x=L_y=140\un{nm}$ for the case {\bf e} or $L_x=L_y=270\un{nm}$ for the case {\bf d}. The periodical boundary conditions are imposed along $x-$ and $y-$axis. Then, the  film is thermalized with the given radial temperature profile $T(r)$ with the fixed $T(r_c)=1500$K. After melting and equilibration the film gains the vertical/separation velocity $\nu_{cm}(r)\sim \cos(\pi r/2r_c)$ with the maximum at the center, presented in Table 1, and zero velocity at the $r_c.$ Because the maximal temperature of the film is much less than the boiling temperature ($\approx 3.2$~kK), the saturated vapor pressure is too small to cause any dynamic effect on the moving film. Indeed, evaporation was not even observed in our MD simulations in the given temperature range.

Our MD-MC simulations of the translative flow-assisted redistribution of the molten film have MC heat transfer resulting in the subsequent resolidification of liquid via heat-conduction cooling of the hot cupola-like shell. The thermal diffusivity shown in Table 1 is adjusted by MC electron jumping rate between neighbor atoms as described in \cite{Inogamov:2010:APA}. A heat sink is simulated by using the Langevin thermostat with the typical temperature of 500 K for all atoms with $r>r_c,$ which leads to resolidification of film from its edge with time. This thermostat maintains also the zero mass velocity outside the circle $r_c,$ which keep mechanically the film outside the circle at the substrate. This condition corresponds to the fixed boundary conditions implied at the edge $r_c$ of the cupola.

MD-MC simulation of a realistic micrometer-sized film having very large amount of involved atoms ($\sim 10^9$)
    during a long cooling time of many tens of nanoseconds is not feasible on available computer resources.
 Instead the small-sized films with several millions of atoms were simulated as indicates Table 1.
 To meet the experimental conditions the scaling approach was used as in \cite{Inogamov:2016:Nanoscale}.
 Two non-dimensional parameters, governing evolution of film separating from substrate with velocity $\nu_{cm},$
 are the capillary parameter
\begin{equation}
 V_{\sigma} = \nu_{\sigma}/\nu_{cm}              \label{eq:Vsig}
\end{equation}
 and the thermal parameter
 \begin{equation}
 V_{\kappa} = \nu_{\chi}/\nu_{cm},               \label{eq:Vkap}
 \end{equation}
 where the capillary velocity  $\nu_\sigma$ and the thermal velocity $\nu_{\chi}$ are defined as
\begin{equation}
\nu_\sigma=2\sqrt{\sigma/(\rho d)},  \qquad \nu_{\chi} = \chi/(2 r_{c}). \label{eq:veloc}
\end{equation}
Here $\chi$ is the thermal diffusivity, $\sigma$ -- surface tension, $\rho$ -- initial density, $d$ -- initial thickness of metal film, and $r_{c}$ is the separation radius.
These governing parameters are also listed in Table 1.

\section{\label{sec:results} Regimes of mass redistribution from experiments and simulations}

\begin{figure}
\includegraphics[width=0.9\columnwidth]{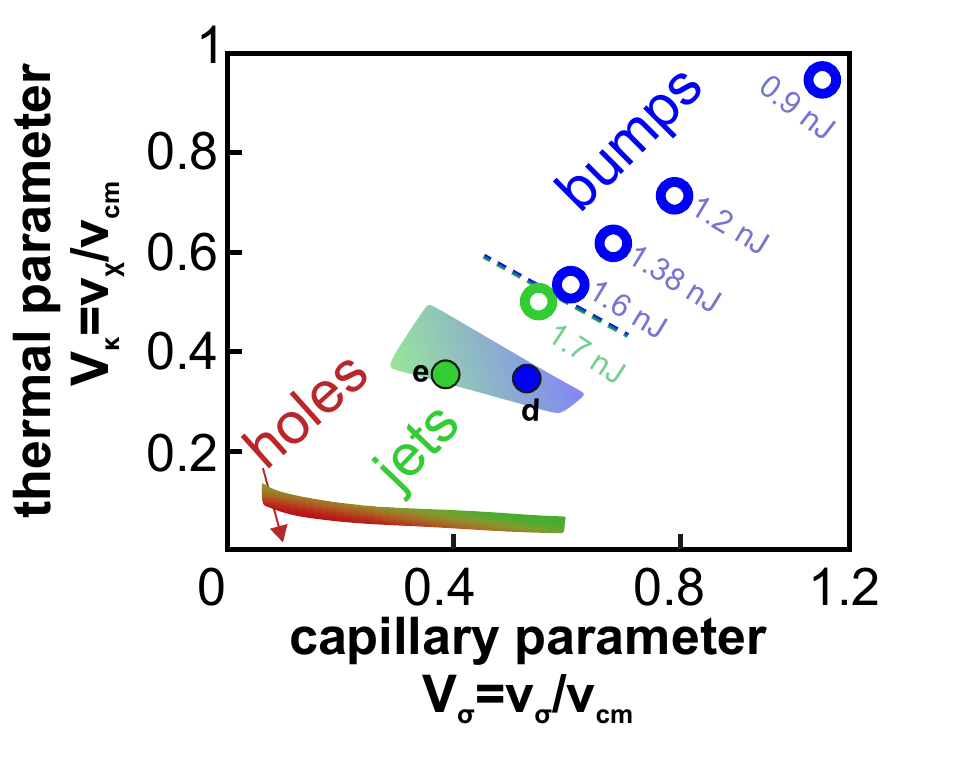}
\caption{(Color online) The map of the thermal $V_\sigma$ and capillary $V_\kappa$ parameters.
 The filled markers "d" and "e" indicate the simulated parameters used to obtain the resolidified geometries presented in Figs.~2(d2) and (e2), respectively. Other simulated markers used to accumulate statistics on the corresponding regimes are not shown to avoid overburdening of the figure. Instead, two gradient-color stripes were used to show the characteristic "hole-jet" (red-green) and "jet-bump" (green-blue) transition boundaries provided by the MD-MC simulations. The experimental capillary and thermal parameters marked as hollow markers were estimated considering the separation velocity of $\nu_{cm}=110$ m/s for the shot with energy 1.7 nJ, see text for explanations. The corresponding applied pulse energies (in nJ) are indicated near each hollow marker. The dashed line shows the experimental "jet-bump" regime boundary. The colors of the markers (both filled and hollow) indicate the transition between different energy-dependent regimes providing formation of the bumps (blue) and the jets (green), while the similar color scheme was used to divide the corresponding pulse energy ranges in Fig.~1(h).}
\label{fig:04}
\end{figure}


 Analysis of the FIB cuts performed on surface features produced at the different incident pulse energies (Fig.~2(a-e))
   indicates several steps, characterizing the molten film evolution
    via corresponding translative hydrodynamic flows and affecting the resulting thickness distribution
      along the produced surface structure:
  (i) inflation of the metal film in the form of the parabolic-shaped cupola, having constant thickness along its circumference;
  (ii) weak redistribution of the cupola thickness,
  (iii) transformation from parabolic to conical shape and, finally, (iv) appearance of the central protrusion.


 Despite the downscaled computation dimensions used in MD-MC simulations, all these steps are perfectly reproduced
   in our theoretical model by varying the thermal and capillary parameters (Fig. 2(b2-e2)),
     which both depend on the incident pulse energy $E,$ in their turn.
 Here, to illustrate the film thickness distribution for parabola-shaped bumps (Fig. 2(b2,c2))
   a few simulated snapshots were taken at different times, when the moving metal film is not completely frozen
    (red and green colors divide the molten and resolidified states of the film).
 By increasing the heat conductivity in the MD-MC simulations, the completely resolidified parabola-shaped bumps
  with the negligible redistribution of mass along the film can be obtained.
 In the case when redistribution of mass is weak the final thickness is approximately equal to the initial thickness.
 This conclusion is in agreement with our previous 2D MD-MC simulations showing weakening of the redistribution
    when absorbed energy is small and conduction cooling is fast \cite{Inogamov:2015:JETP,Inogamov:2016:JPCC:Blis}.
 The frozen shapes are similar for 2D and 3D geometries in case of fast cooling. Upscaling of the film thickness via thin-shell recalculation \cite{Inogamov:2016:JPCC:Blis}
   also gives perfect qualitative agreement with the experimentally measured thickness distributions for all surface structures (Fig. 2(a1-e1)), providing the reliable way for theoretical simulation of the fabricated surface 3D-features.


 Other common surface structures, such as high-aspect-ratio nanojets and through holes (Fig.~1(f-h)), can be also reproduced by varying the capillary and thermal parameters in the MD-MC simulation. Carrying out such simulations as well as using our recently published results \cite{Inogamov:2016:Nanoscale}, we have built the map of the corresponding non-dimensional governing parameters required to reproduce the certain type of the structure, where the transitions (or boundaries) between "bump-jet" and "jet-hole" regimes are marked with the transition blue-green and green-red strips, respectively (see Fig.~4). Specifically, two important dots marked as "d" and "e" letters, reflecting the parameter set for MD-MC simulations from the Figs.~2 (d2) and (e2) as well as establishing theoretical "bump-jet" regime boundary, are indicated in this map. These two important regimes can be formally divided according to relative translative-flow assisted mass redistribution of the molten film near the bump tip. This separation follows from the analysis of the simulated structures with moderate and significant material redistribution near the structure central axis (Figs.~2(d2) and (e2), respectively). Similar formal separation can be also applied for our experimental results (see Fig.~1(d), (e) and (h) and discussions below). The corresponding parameter set for "d" and "e" dots are summarized in the Table 1.





 We estimate the thermal and capillary parameters, which correspond to our five main experimental points (see Fig. 2(a1-e1)) on the ($V_\sigma$,$V_\kappa$) parameter plane (hollow circles in Fig.~4) using the typical diameter of the bump $2r_c\approx 900$nm from Fig.~1(h) and corresponding Eqs.(\ref{eq:Vsig}-\ref{eq:veloc}). Taking into account the temperature-dependent surface tension of molten gold $\approx 0.9\un{J/m^2}$ \cite{Inogamov:2015:JETP} and the thermal diffusivity $\chi \approx 0.5 \un{cm^2/s},$ for our 50-nm thick Au film the thermal and capillary velocities are equal to $\nu_\chi\approx 56 \un{m/s}$ and $\nu_\sigma\approx 61 \un{m/s}$, respectively. Although the experimental values of $\sigma,$ $\kappa,$ $r_c$ and the corresponding velocities $\nu_\sigma$ and $\nu_\kappa$ are well defined, the experimental separation velocity $\nu_{cm}$, required to estimate the $V_\sigma$ and $V_\kappa$, is an unknown parameter.


 For the fixed diameter of the optical spot and the absorption coefficient $A$ weakly dependent on the incident fluence $F_{inc},$
  which holds true in our energy range, the absorbed fluence $F_{abs}$ will be proportional to the pulse energy $E.$
 From the other hand, according to our 2T-HD simulations of the separation stage
   and considering negligibly weak adhesion between the Au film and the silica substrate,
    the velocity $\nu_{cm}$ is proportional to the applied pulse energy $E$ \cite{Inogamov:2016:JPCC:Blis}.
 Using this proportion and assuming that
 \begin{equation}
 \label{eq:NUcmProptoEpulse}
 \nu_{cm}(E) = (\nu_{cm}|_{E=1.7})*(E/1.7),
 \end{equation}
 where $E$ is given in nJ and $\nu_{cm}|_{E=1.7}$ is an adjustable separation velocity at $E=1.7$ nJ,
  one can obtain the set of the experimental governing parameters on the capillary-thermal plane
   $(V_\sigma,V_\kappa)$ (\ref{eq:Vsig},\ref{eq:Vkap})
      related to our particular experimental cases for the fixed $\nu_{cm}|_{E=1.7}$ value.



 To obtain the separation velocity $\nu_{cm}$ as a function of the absorbed fluence $F_{abs}$ and assess its reliable value, the one-dimensional 2T-HD modeling of film on glass substrate was performed (see Supplemental material for details). Assuming the $\nu_{cm}|_{E=1.7}$= 110 m/s and using Eq. (\ref{eq:NUcmProptoEpulse}), we obtain the set of governing parameters on the $(V_\sigma$,$V_\kappa)$ plane (hollow markers in the Fig. 4), which corresponds to our experimental results (Figs. 2(a1-e1)).
Analysis of the parameter plane indicates that the experimental ``bump-jet'' transition boundary (dashed curve in the Fig. 4) defined by the surface structures obtained at pulse energies E=1.6 and 1.7 nJ (blue and green markers, respectively) is located 35\% higher than the corresponding boundary estimated from the analysis of the simulation results. This deviation can be explained by (i) the difference between the boundary conditions in simulation and experiment, as well as (ii) by the reinforcement of a film to the internal rupture, which shifts the threshold of internal rupture $F_{rup}$ to larger magnitudes of absorbed fluence $F$ thanks to 3D geometry (see discussions in the Supplemental material).


 Finally, it is plausible that the relative ratio between the ``bump-jet'' and ``jet-hole'' threshold boundaries is weakly sensitive to the difference in the mentioned boundary conditions. The experimentally measured ratio of the characteristic threshold energies required for the formation of the through hole and the jet (corresponding energy logarithms $ln(E)$ are equal to 1.9 and 0.65 in Fig. 1(h)) is about 3.5. Analysis of the calculated capillary-thermal parameter map (position of the gradient-color stripes in Fig.~4) gives the ratio equal to 3.6. This ratio is calculated measuring the distances from the origin of the two characteristic dots on the intersection of the diagonal line in Fig.~4 and the theoretical `bump-jet'' and ``jet-hole'' threshold boundaries.  Thus, the theoretical and experimental results are in satisfactorily good  agreement with each other.

\section{\label{sec:level1}Conclusions and Outlook}

Highly irreversible transient nanoscale translative hydrodynamic flows, quenched via nanoscale heat conduction, formed  radially-symmetric topological surface patterns: nano-bumps and jets. They were made by single-shot irradiation of a 50-nm-thick gold film by tightly focused femtosecond laser pulses at variable energies. Radial mass distribution (thickness) across the individual features was \emph{quantitatively} acquired either by their nanoscale focused-ion beam cutting and following side-view electron microscopy analysis, or by direct top-view radial nanoprofiling of their relative thickness using energy-dispersive x-ray fluorescence spectroscopy at different electron-beam energies, calibrated by the signal from the unperturbed film. Molecular dynamics simulations were undertaken to envision spatiotemporal dynamics of the underlying nanoscale hydrodynamic melt flows and were shown to provide \emph{quantitative} predictions of the evolution of the metal film thickness.

More generally, on the one hand, this study demonstrates that highly-transient nanoscale hydrodynamic melt displacements and their dynamics can be precisely modeled, enabling in future realistic predictions of more complex 3D nanostructures and 3D structure materials ~\cite{16le16133} produced by focused ultra-short laser pulses, carrying, for example, angular momentum~\cite{Toyoda2013} ("vortex laser pulses"), and accurate simulations of more complicated systems, such as materials with new metastable amorphous and crystalline phases~\cite{11nc445,Rapp2015}, phase-change materials \cite{Hase2015,Matsubara2016} or metallic glasses~\cite{Zhong2014}. On the other hand, our quantitative characterization of the mass distribution profiles for the laser-fabricated 3D nanoscale structures lays down, for the first time, a solid basis for solution of the reverse problem, related to their evolution dynamics, to benchmark theoretical models. We can envisage that corrections required to equation-of-state to predict modifications of material subjected to the high pressure and temperature can be established.

\section{\label{sec:appndx} Supplemental material}

\begin{figure*}
\includegraphics[width=0.7\textwidth]{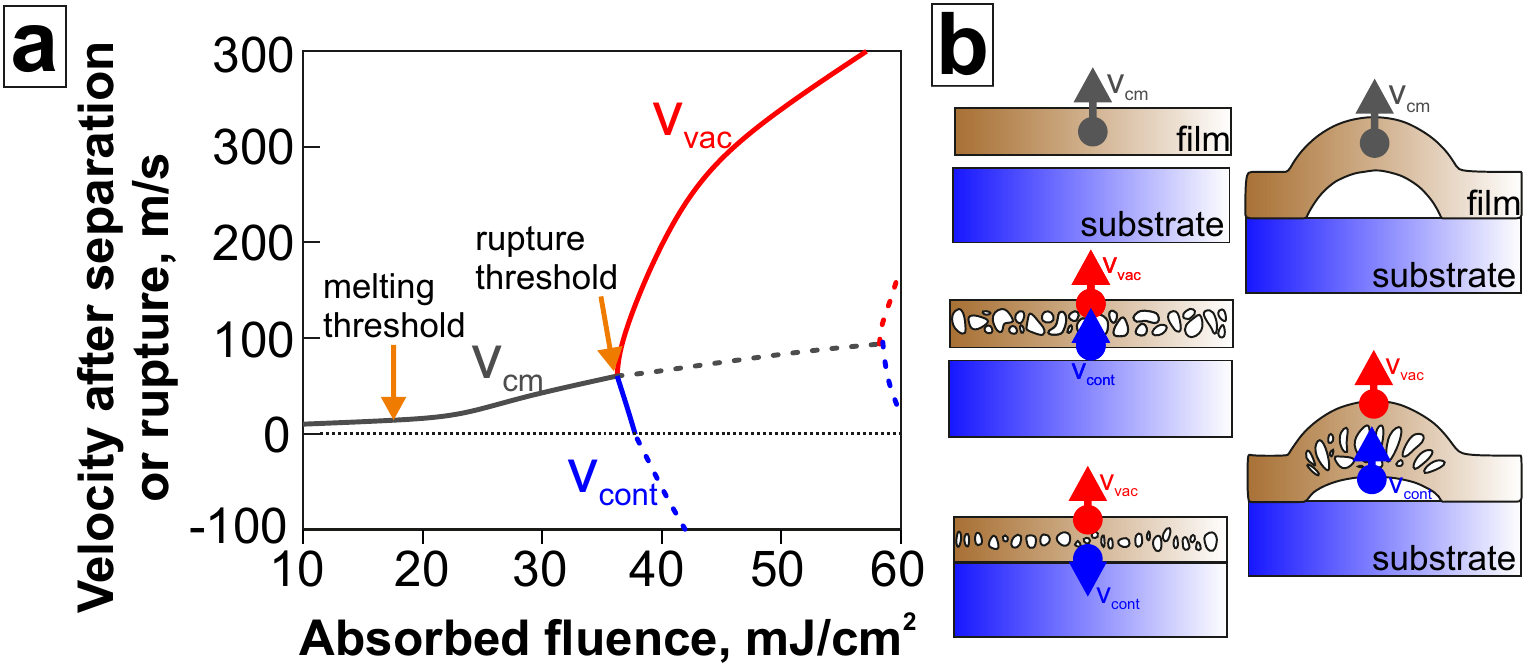}
 \caption{(Color online) (a) Velocity $\nu_{cm}$ of the separated metal film
  as a function of the absorbed fluence $F_{abs}.$
 There are two different fluence-dependent separation regimes schematically shown in the left column of the right Figure (b).
 At the fluences $F_{abs}$ lower than the rupture threshold $F_{rup}$ the film separates as a whole
   having the local center-of-mass velocity $\nu_{cm}(F_{abs})$ (see the upper picture in the left column in (b)),
    while at higher fluences the film splits into "vacuum" and "contact" parts
     having averaged velocities $\nu_{vac}(F_{abs})$ and $\nu_{cont}(F_{abs}),$ respectively;
      see the middle and bottom pictures in the left column in (b).}
\label{fig:05}
\end{figure*}


 Because the experimental separation velocity was adjusted
  to provide the agreement between experimental and simulation results,
    additional discussion is given here to substantiate the chosen separation velocity
$$
\nu_{cm}|_{E=1.7} = 110\un{m/s}.
$$


 To estimate the dependence of the $\nu_{cm}$ on the absorbed fluence $F_{abs},$
  we have carried out the one-dimensional 2T-HD simulations
   neglecting adhesion of gold to the silica substrate thanks to hydrophobic property of silica against molten gold.
 The 2T-HD simulations show that the film is homogeneously heated across its thickness $d$   
  during $\approx 1 \un{ps}$ after the ultrashort pulse irradiation \cite{Inogamov:2016:JPCC:Blis}.
 Fast spreading of absorbed energy from a skin-layer proceeds due to an electron thermal conductivity
  strongly enhanced at the 2T stage.
 According to our simulations (see Fig.~5(a)), below the threshold $F_{rup}$ of the internal rupture of a film,
   a film separates as a whole from substrate collecting the velocity $\nu_{cm}(F_{abs})$
    during the temporal interval when a film remains in mechanical contact with substrate.
 Velocity $\nu_{cm}(F_{abs})$ is obtained after the instant of separation by averaging of local velocities over the thickness of a film.
 Thus it is velocity $\nu_{cm}$ of a center of mass (c.m.) of a film.


 Two rarefaction waves V and C running towards each other
   from the "vacuum-film" (V - vacuum) and "substrate-film" (C - contact) boundaries
    create a layer with negative pressure $p_{tensl}$ near the middle plane of a film,
     see details in \cite{Inogamov:2015:JETP,Inogamov:2016:JPCC:Blis}.
 These two waves begin to interact at acoustic timescale $t\approx t_s/2;$
   $t_s=d/c_s$ $d$ is thickness of a film, $c_s$ is a sound speed.
 For fluences $F<F_{rup}$ the tensile stress $|p_{tensl}|$ created by V-C interaction
   is below the material strength $p_{strength}$ of gold and the film survives (no rupture) under stretching.


 If the film survives then the wave V running from the vacuum side continues its run in direction to the contact
  (the boundary between a film and substrate).
 The wave V arrives at the contact at the instant $t\approx t_s.$
 Pressure $p_c(t)$ at the contact evolves in time.
 Acoustic impedance of gold $Z_g$ is approximately 7 times larger than impedance of silica $Z_s.$
 Pressure $p_c$ is significantly less than the pressures $p$ in the bulk of a film: $p_c\approx (Z_s/Z_g) p.$
 When the wave V achieves the contact then the function $p_c(t)$ changes its sign from positive to negative \cite{Inogamov:2015:JETP,Inogamov:2016:JPCC:Blis}.
 Adhesion is small therefore the separation of a film as a whole from substrate
   (shown in the upper picture in the left column in Fig.~5(b))
    begins around the time $t\approx t_s.$ The separation is caused by the tensile action of the wave V onto the contact.



 Situation changes dramatically if $F > F_{rup},$ the tensile stress $|p_{tensl}|$ overcomes the strength $p_{strength},$
   and the film begins to split into the vacuum and the contact parts {\it before} rupture of the contact.
 This is shown in the middle and bottom pictures in the left column in Fig.~5(b).
 Velocities $\nu_{vac}$ and $\nu_{cont}$ in these two pictures refer to the vacuum and the contact parts of a film
   at the stage when the parts lose their mechanical connections.
 Values $\nu_{vac}$ and $\nu_{cont}$ are obtained by averaging along thicknesses of the vacuum and the contact parts.



 As was said, the ratio of impedances of gold $Z_g$ to silica $Z_s$ is large.
 Therefore expansion of a film after action of an ultrashort laser pulse is approximately symmetric
   relative to the middle plane of a film.
 Small deviation from symmetry is caused by small but finite ratio of impedances $Z_s/Z_g$
  (and also because $d<d_T,$ but not $d\ll d_T,$ where $d_T$ is thickness of a heat affected zone).
 In the case when $Z_s/Z_g=0$ and film is thin (thus it is almost homogeneously heated across its thickness),
   velocity $\nu_{cm}$ is $\approx 0$ in spite of significant expansion velocities.
 The expansion velocities connected with the rarefaction waves V and C (introduced above) are opposite in their directions
   and compensate each other
      giving resulting small velocity of the center of mass $\nu_{cm}(F_{abs})$ if $F<F_{rup},$ see Fig.~5(a).


 Above the rupture threshold $F_{rup}$ the center of mass velocities $\nu_{vac}$ and $\nu_{cont}$
  of the vacuum and the contact parts are split;
   for $F < F_{rup}$ we can formally write $\nu_{cm}=\nu_{vac}+\nu_{cont}.$
 There is a short interval near the $F_{rup}$ in Fig.~5(a) where velocities $\nu_{vac}$ and $\nu_{cont}$
   have the same sign directed into the vacuum side.
 The situation corresponding to this interval is illustrated by the middle scheme in the left column of Fig.~5(b).


 Above this interval the contact part moves in direction to the substrate.
 The continuation of the $\nu_{cont}(F_{abs})$ âóçóòâóòñó into the range of negative velocities $\nu_{cont}<0$
 is shown by the bottom dashed curve in Fig.~5(a). Of course, this is the conventional curve
 because the magnitudes $\nu_{cont}(F_{abs}) < 0$ are not asymptotic magnitudes $t\to \infty$ conserved in time.
 2T-HD simulations and analytic analysis show that dense gold (with high $Z_g)$ from the contact part
   presses the less dense silica (with low $Z_s)$
    supporting positive contact pressure $p_c(t)$
     exponentially decaying in time $p_c(t)\propto \exp(-t/t_{decay})$
       with e-folding time $t_{decay}$ of the order of $t_s.$
 This pressure decelerates motion of the contact part and thus decreases velocity $\nu_{cont}(t).$


 We are sure that at the acoustic time scale $\sim t_s$ the velocity of the contact part remains directed to the substrate side
   (no separation from substrate).
 The fate of the contact part of the gold film at the late stage $t\gg t_s$ is unclear.
 At least in our experiments with extremely small irradiation spot $r_c=300-400$ nm described in this paper (see Fig.~2(a1-e1))
   we don't see remnants of gold
    on the surface of silica inside the cavity under the shell of the bump.
 Maybe the liquid gold film fragments into a set of droplets on a surface of hydrophobic silica,
  and the droplets move to the periphery of the circle under the shell
   under action of 3D effects and velocities directed along the surface;
     also partial evaporation and re-deposition onto the rear-side of the shell takes place.


 Near the threshold $F > F_{rup}$ the velocity $\nu_{vac}$ quickly increases with difference $F-F_{rup}>0,$ see Fig.~5(a).
 Below the threshold $F < F_{rup}$ the momentum
 \begin{equation}
 \label{eq:03-momentumVacuumPart}
 I = \int_{-\infty}^{ts/2} p(x=d/2, t) dt
 \end{equation}
  expanding the vacuum part of a film
   is compensated by resistance of condensed gold to stretching, here $p(x=d/2,t)$ is pressure in the middle of the film.
 If we overcome the threshold $F_{rup}$ to a few tens of percents then the counter-momentum produced by the resistance force of condensed matter becomes small relative to the momentum (\ref{eq:03-momentumVacuumPart}).
 Then the center of mass velocity of the vacuum part is $\nu_{vac}\approx I/(\rho\, d/2).$
 This velocity roughly is $Z_g/Z_s\approx 7$ times larger than velocity $\nu_{cm}$ near threshold.
 This explains the growth of $\nu_{vac}(F_{abs})$ above threshold $F_{rup}$ in Fig.~5(a).



 According to the particular series of the 1D 2T-HD simulations gathered in Fig.~5(a), the values $F_{abs}|_{rup},$ $\nu_{cm}|_{rup}$
   describing the threshold of rupture
     are rather small: $F_{abs}|_{rup}\approx 40$ mJ/cm$\!^2,$ $\nu_{cm}|_{rup}\approx 60$ m/s.
 The large ratio $Z_g/Z_s$ results in decreasing velocity $\nu_{cm}$
   and increasing susceptibility of velocity $\nu_{vac}(F_{abs})$ to the raising of energy $F_{abs}$
    above the threshold $F_{rup},$ see Fig.~5(a).
 Why do we say that the values $F_{abs}|_{rup}$ and $\nu_{cm}|_{rup}$ are small? Explanations are given below.
 The main problem is connected with the comparison of relative positions of the rupture threshold $F_{rup}$ from Fig.~5 and the threshold $F_{hole}$ from Fig.~4 for appearance of a through hole; the positions on the energy axis in Fig.~5. Fig.~5 presents the fast stage, it doesn't know about capillary forces, $r_c,$ and the rate of slow cooling at the one-temperature stage. Definitely, $F_{hole}<F_{rup}$ for large spots $r_c.$ But for small spots the situation is complicated as we see below.


 Let $\nu_\sigma\approx \nu_\chi$ (\ref{eq:veloc}), then we are near the bisectrix of the angle
   between the horizontal and vertical axes in Fig.~4.
 The bisectrix cuts the "bumps-jets" and "jets-holes" boundaries at the points
   $V_\sigma=V_\kappa\approx 0.38$ and $\approx 0.1$ respectively.


 In our experiments the boundary conditions are softer than in MD-MC simulation:
   only at the infinity a film is held on the substrate (adhesion is weak)
    and only at the infinity the temperature returns to the initial temperature of a film equal to 300 K.
 While in MD-MC simulation a film is held on the substrate in the narrow rim around the radius $r_c;$
  and cooling of the inflated shell is carried on by supporting fixed temperature below melting temperature in this rim, see Section \ref{sec:methods}.


 If we use the softer conditions in simulations then the threshold velocity $\nu_{cm}$
   necessary to shift from the regime with bumps to the regime with jets will be smaller,
    because the cupola is weaker mechanically connected to the film
     and cooling rate is lower since the cold bath/sink is more distant.
 The estimates give 30-40\% decrease of the threshold velocity $\nu_{cm}.$
 Thus the intersections of the bisectrix with the "bump-jet" and "jet-hole" boundaries
   in conditions more close to the experiment will be in the points
      $V_\sigma=V_\kappa\sim 0.54$ and $\sim 0.14$ respectively.



 In our experiments $v_\sigma\approx 60$ m/s, $v_\chi\approx 60$ m/s (\ref{eq:veloc}), see Section \ref{sec:results}.
 If we take the threshold values $V_\sigma=V_\kappa\sim 0.54$ and $\sim 0.14$ corrected above
   then the bumps exist in the range of velocities
      $0 < \nu_{cm} < 110$ m/s,
       while the jets and holes appear when $110 < \nu_{cm} < 400$ m/s and $\nu_{cm} > 400$ m/s, respectively.
 Comparing these high velocities with those in Fig.~5(a), we see that the bumps, jets, and through holes
  fall into the set of cases above the rupture threshold.



 The rupture threshold $F_{rup}$ in Fig.~5(a) has been found from the 1D 2T-HD runs.
 The value $F_{rup}$ depends on the model of nucleation used in the code.
 The model has been presented in \cite{Khokhlov:2016:JPCC}.
 The model was adjusted to the spallation results of MD-MC simulations \cite{Demaske:2010}
   and may somewhat underestimate the material strength of gold.
 If we calculate energy density $\epsilon_{rup} = F_{rup}/d$
  before the bulk expansion stage
   at the threshold $F_{rup}=36 \un{mJ/cm^2}$ for $d=60$ nm
    as in Fig.~5(a) then we will obtain $\epsilon_{rup}\approx 0.64$ eV/atom.
 The heat of vaporization (cohesive energy $E_{coh})$ of gold is 342 kJ/mol = 3.4 eV/atom.
 According to paper \cite{Upadhyay:2008} the ablation threshold $F_{abl}$ for a freestanding Al or Cu film
  is 28-30\% of $E_{coh};$
   above the threshold $F_{abl}$ the freestanding film separates into two halves;
   thus the ablation threshold is equivalent to the rupture threshold above which a film also separates into two parts.
 Cu is similar to gold, therefore $\epsilon_{rup} = 0.95-1$ eV/atom according to \cite{Upadhyay:2008}.



 The shift from $\epsilon_{rup}\approx 0.64$ to $0.95-1$ eV/atom increases $F_{rup}$ and $\nu_{cm}|_{rup}$
   to 60 mJ/cm$\!^2$ and 90 m/s.
 Corresponding continuation of the regime of separation as a whole
 with dependence $\nu_{cm}(F_{abs})$
 and the shift of the threshold $F_{rup}$
  are shown in Fig.~5(a) by the dashed-line curves.
 Additional corrections are related (i) to the  small decrease of the threshold $\epsilon_{rup}$
  from the value $0.95-1$ eV/atom
   because our film is 2-3 times thicker than the films studied in \cite{Upadhyay:2008};
   (ii) and with the small increase (from the value $0.95-1$ eV/atom) proportional to $Z_s/Z_g$
    since due to presence of a substrate
     the rarefaction wave C coming to the middle of the film from the substrate side
      is weaker than the rarefaction produced by expansion into vacuum
       (substrate becomes mechanically equivalent to vacuum in the limit $Z_s/Z_g\to 0).$



 Additional factors increasing the threshold $F_{rup}$ in Fig.~5(a) are linked to
  (i) increase of resistance to stretching thanks to existence of foam
   and to (ii) 3D dimensional character of blistering (see the right pictures in Fig.~5(b)).
 These factors are illustrated in Fig.~5(b).
 The middle and bottom pictures in the left column in Fig.~5(b) schematically present membranes in the foam
  in the plane case: $r_c=\infty.$
 The membranes mechanically link the vacuum and contact parts of the film after nucleation.
 These links reinforce a film against rupture.


 In the model of rupture used in the 1D 2T-HD code, the condensed matter is ruptured (pressure drops to zero)
   as the criterium of nucleation is fulfilled.
 Of course, this is an approximation. The real situation is shown in Fig.~5(b).
 The foam and the curvature of the bump hamper the total rupture,
  thus increasing the values $F_{rup}$ and $\nu_{cm}|_{rup}.$
 Nevertheless the situation with blistering under the above-threshold conditions needs additional studies.



 The analysis presented above shows that even in the range of high fluences $F_{abs}=40-80 \un{mJ/cm^2}$
  still it is possible
   to keep the bump with the jet structures at the surface of substrate
    (avoiding its total rupture and formation of a through hole)
      in spite of high values of velocity $\nu_{vac}.$
 This is possible due to the high aperture objective and the extremely small (relative to all previous experiments) size
  of an irradiated spot $r_c=300-400$ nm, see Fig.~2.



 Velocity $\nu_{vac}$ is established after the breaking of the mechanical connection through the foam
   between the vacuum and contact parts of the film.
 The only slow mechanical deceleration by surface tension directed tangentially along the surface of the vacuum part
   remains after rupture of the foam.
 This kind of deceleration appears due to ties of the flying vacuum part with the solid film
   outside the separation radius.
 This deceleration becomes more significant with increase of curvature of the flying film.
 Until the foam survives the curvature is small and deceleration of the vacuum part
   is accomplished by the liquid membranes forming the foam;
    this is illustrated in the left column in Fig.~5(b).
 At this stage the forces are directed normally to the surface of the vacuum part.



 It should be mentioned again that below the rupture threshold $F_{rup}$ the film separates as a whole.
 Thus the rear-side (opposite to the vacuum side) of the shell of the bump is smooth
   (see the upper pictures in the left and right columns in Fig.~5(b))
    because the contact between a film and substrate is smooth
      and separation (the rupture of a contact) of a film from substrate is continuous (no random nucleations).
 Above threshold $F_{rup}$ the internal rupture of a film takes place.
 This process begins with random nucleations in the middle of the film (see the bottom pictures in Fig.~5(b)).
 The process continues with formation of two-phase liquid-vapor mixture transforming gradually into foam
   between the vacuum and contact parts of the film separated later from each other.
 This stage is shown in the bottom pictures in Fig.~5(b).

 As was said above, the liquid membranes inside the foam
   connect for some time mechanically and thermally the vacuum and contact parts.
 Due to mechanical links the vacuum part is slightly decelerated during this transition stage
   (transition from the connected stage to the separated stage).
 The thermal connections cool the vacuum part through foam.
 The remnants of the foam exist for some time at the rear-side of the vacuum part.
 Thus above threshold $F_{rup}$ for some time the rear-side cannot be regarded as a smooth surface
   opposite to the cases below threshold $F_{rup};$ compare the up and bottom pictures in the right column in Fig.~5(b).


 High values of velocity $\nu_{vac}$ mean that the first droplet will also have high velocity.
 This droplet appears above the threshold for separation of the first droplet from the jet.
 The first droplets fly perpendicularly to the surface of substrate with very narrow angular distribution.
 Thus using of a high-aperture lens and focusing into extremely small spot is the way to production
  of such fast, exactly directed droplets.
 This is important in technologies of printing like laser-induced forward/backward transfer (LIFT/LIBT).

{\begin{acknowledgments}
The experimental part of this research was supported by the Russian Science Foundation (grant no. 17-19-01325). S.J. acknowledges Workshop of Photonics R\&D. Ltd. for the laser fabrication setup acquired via a collaborative grant and the Australian Research Council DP170100131 Discovery project. V.V.Z. and N.A.I. are responsible for theoretical part of the paper, simulations, and comparison with experimental data; they acknowledge grant of the Russian Science Foundation (project no. 14-19-01599).

\end{acknowledgments}}

\end{document}